\documentclass[12pt]{article}
\usepackage[dvips]{graphicx}
\usepackage{epsfig,cite}
\usepackage {amsmath}
\usepackage{color}
\usepackage{amssymb}
\usepackage{slashed}
\include{epsf}
\newcount\timecount
\newcount\hours \newcount\minutes  \newcount\temp \newcount\pmhours
\hours = \time
\divide\hours by 60
\temp = \hours
\multiply\temp by 60
\minutes = \time
\advance\minutes by -\temp
\def\hour{\the\hours}
\def\minute{\ifnum\minutes<10 0\the\minutes
            \else\the\minutes\fi}
\def\clock{
\ifnum\hours=0 12:\minute\ AM
\else\ifnum\hours<12 \hour:\minute\ AM
      \else\ifnum\hours=12 12:\minute\ PM
            \else\ifnum\hours>12
                 \pmhours=\hours
                 \advance\pmhours by -12
                 \the\pmhours:\minute\ PM
                 \fi
            \fi
      \fi
\fi
}

\def\monthname{\relax\ifcase\month 0/\or January\or February\or
   March\or April\or May\or June\or July\or August\or September\or
   October\or November\or December\else\number\month/\fi}

\def\bold#1{\setbox0=\hbox{$#1$}%
     \kern-.025em\copy0\kern-\wd0
     \kern.05em\copy0\kern-\wd0
     \kern-.025em\raise.0433em\box0 }

\newcommand{\be}{\begin{equation}}
\newcommand{\ee}{\end{equation}}
\newcommand{\bear}{\begin{eqnarray}}
\newcommand{\eear}{\end{eqnarray}}

\newcommand{\vev}[1]{\left\langle #1\right\rangle}

\newcommand{\GeV}{\; \mathrm{GeV}} 
\newcommand{\TeV}{\; \mathrm{TeV}} 
\newcommand{\lapproxeq}{\lower .7ex\hbox{$\;\stackrel{\textstyle  
<}{\sim}\;$}} 
\newcommand{\gapproxeq}{\lower .7ex\hbox{$\;\stackrel{\textstyle  
>}{\sim}\;$}} 
\newcommand{\stackdown}[2]{\lower 1.4ex\hbox{$\;\stackrel{\textstyle{#1}}  
{\scriptstyle{#2}}\;$}}
\newcommand{\beq}{\begin{equation}} 
\newcommand{\eeq}{\end{equation}} 
\newcommand{\ba}{\begin{eqnarray}}
\newcommand{\ea}{\end{eqnarray}}

\newcommand{\bea}{\begin{eqnarray}}
\newcommand{\eea}{\end{eqnarray}}

\def\gappeq{\mathrel{\rlap {\raise.5ex\hbox{$>$}}
{\lower.5ex\hbox{$\sim$}}}}
\def\lappeq{\mathrel{\rlap{\raise.5ex\hbox{$<$}}
{\lower.5ex\hbox{$\sim$}}}}
\def\Toprel#1\over#2{\mathrel{\mathop{#2}\limits^{#1}}}

\makeatletter 
\def\slash{\@ifnextchar[{\fmsl@sh}{\fmsl@sh[0mu]}} 
\def\fmsl@sh[#1]#2{% 
  \mathchoice 
    {\@fmsl@sh\displaystyle{#1}{#2}}% 
    {\@fmsl@sh\textstyle{#1}{#2}}% 
    {\@fmsl@sh\scriptstyle{#1}{#2}}% 
    {\@fmsl@sh\scriptscriptstyle{#1}{#2}}} 
\def\@fmsl@sh#1#2#3{\m@th\ooalign{$\hfil#1\mkern#2/\hfil$\crcr$#1#3$}} 
\makeatother 
\begin{document}

\begin{titlepage}
\pagestyle{empty}
\baselineskip=21pt
\vskip 0.2in
\begin{center}
{\large {\bf The Price of a Dark Matter  Annihilation\\ Interpretation of   AMS-02 Data}}

\end{center}
\begin{center}
\vskip 0.5in
 {\bf Vassilis~C.~Spanos} 
\vskip 0.1in
{\small {\it
{Institute of Nuclear \& Particle Physics, NCSR ``Demokritos'', GR-15310 Athens, Greece}} \\
}
\vskip 4cm
{\bf Abstract}
\end{center}
\baselineskip=18pt \noindent
%%%%%%%%%%%%%%%%%%%%%%%%%%%%%%%%%%%%%%%%%%%%%%%%%
{\small

We discuss challenges to a dark matter annihilation interpretation of the excess positron fraction
in the cosmic rays observed by the PAMELA, Fermi-LAT and AMS-02 collaborations.
The spectra of positrons from annihilations into the
leptonic two-body final states like $\tau^+ \tau^-$ or $\mu^+ \mu^-$,
 fit well the AMS-02 data for the positron ratio and the electron  flux.
Furthermore, we discuss the hadronic annihilation channels  $b \bar{b}$ and  $W^+ W^-$.
 However, this interpretation requires a very large annihilation
cross section especially for the hadronic channels,  conflicting with the unitarity upper limit, unless the
positron flux due to annihilations is boosted by a large factor due to
inhomogeneities in the galactic halo.
 In addition, we present predictions within this interpretation for the positron fraction at higher energies
for the antiproton flux, and discuss constraints from $\gamma$-ray
measurements. }

%%%%%%%%%%%%%%%%%%%%%%%%%%%%%%%%%%%%%%%%%%%%%%%%

\vfill
\leftline{December 2013}
\end{titlepage}
\baselineskip=18pt
%%%%%%%%%%%%%%%%%%%%%%%%%%%%%%%%%%%%%%%%%%%%%%%%%%

\section{Introduction}

The recent precise data from the AMS-02 experiment~\cite{AMS02} have rekindled
speculation about the origin of the cosmic-ray positron excess
observed previously by the PAMELA~\cite{pamela_rat} and Fermi-LAT~\cite{Fermi_rat} collaborations.
Similar data   have been reported earlier  by    the HEAT~\cite{HEAT} and AMS-01~\cite{AMS1} experiments.
The AMS-02 positron ratio data are much more precise than those from PAMELA and  Fermi-LAT,
and agree better with the former. In addition, they are extending to higher energies.
On the other hand, the Fermi-LAT
data for the positron ratio (fraction) are less precise and agree less well with PAMELA, thus they  will  not be used in our analysis.
Therefore,  in the following, we discuss mainly fits to AMS-02 positron ratio data.

The most conservative interpretation of the positron excess is based
on local astrophysical sources such as nearby pulsars~\cite{Yin:2013vaa,Linden:2013mqa,Cholis:2013psa}
or secondary effects~\cite{Blum:2013zsa}, whereas less
conservative interpretations are based on the annihilations or
decays of massive cold dark matter (DM)~\cite{EHNOS} particles~\cite{back_int,Yuan:2013eja,Jin:2013nta,Dienes:2013lxa,Li:2013poa}. 
The positron spectra produced by local sources cannot be predicted
precisely, but seem capable of fitting the positron excesses observed
by PAMELA, Fermi-LAT and AMS-02. The `smoking gun' for an
astrophysical interpretation would be an isotropy in the excess positron flux, but
the current sensitivity of the AMS experiment is not enough to
test this possibility. The positron spectra produced in DM
annihilations depend on the final states, but precise
predictions can be made once a specific final state is assumed.
Moreover, once the final state is specified, definite predictions can be made
for the spectra of other annihilation products such as $\gamma$-rays
and antiprotons.

Many theoretical papers have explored potential DM annihilation interpretations
of the positron excess, both before and after the recent AMS-02 data release.
Our focus here is primarily on challenges that such interpretations must
overcome. There is, however, another issue concerning  the required annihilation cross section,
namely that it may exceed the unitarity limit $(4 \pi / m_\mathrm{DM}^2) \times (2 J + 1)$~\cite{Griest:1989wd}
for plausible values of the initial state angular momentum, $J = 0, 1$. As we
see later, this constraint would be violated over much of the DM mass range,  that 
would otherwise be a favored fit to the AMS data. In particular, only the DM annihilation channel
 $\mu^+\mu^-$, and partly then $\tau^+\tau^-$,  survive the unitarily bound.

This challenge may be surmounted by postulating a suitable boost factor $B$
originating from inhomogeneities in the density of DM in the galactic
halo, e.g., a boost factor $B \sim 100$ would be required to respect the
unitarity constraint if $m_\mathrm{DM} \sim 10$~TeV. A further boost by a factor
$\gappeq 100$ would be needed if the cosmological density of DM
particles had an exclusively thermal origin. We leave it to astrophysicists
to assess whether such large boost factors are plausible, but they do seem
rather large.

If one could overcome  these challenges arising from the DM annihilation rate, 
these  scenarios can be used to make predictions for
the fluxes of antiprotons ${\bar p}$ and $\gamma$-rays. The experimental constraints
on the antiproton flux are not yet sufficient to exclude the leptonic annihilation channels $\mu^+ \mu^-$ and $\tau^+ \tau^-$,
 though future AMS ${\bar p}$ data may provide an
interesting constraint. 
The constraints imposed by available $\gamma$-ray
data are, however, already quite severe. Those provided by the flux from
the galactic centre (GC)  are rather dependent on the halo model, but the
constraints on the $\gamma$ flux from other sources 
like the dwarfs spheroidal galaxies (dSph)~\cite{dsphfermi_o,dsphfermi} 
 are not so  model-dependent and can provide reliable bounds to the DM annihilation models.

 There are two basic approaches to interpret the AMS-02 data using the DM annihilation products. 
 One  is to fit the data and interpret them as the cosmic ray  (CR) background~\cite{back_int}.
 Thus, the room that is left to the DM contribution is minute,
 yielding to masses of the DM particles $m_\mathrm{DM}$ of the order of few hundreds of GeV
and to annihilation cross-section $\vev{\sigma v}$ of the order of $10^{-26}\, \mathrm{cm^3/s}$ or less. 
This results doesn't require big boost factors to reconcile the AMS-02 constraint
with the cosmological DM abundance bound. 
Similarly,  one can   interpret  the AMS-02 positron ratio data  as an effect of secondary reactions~\cite{Blum:2013zsa}.
Unfortunately, it appears that this  interpretation may not be compatible to  the 
 boron to carbon  radioactive nuclei ratio  data (B/C)~\cite{Cholis:2013lwa}.

The second approach, which we have adopted here,    is more conventional. We adjust  the {\tt GALPROP} 
conventional model, that is compatible to B/C data,
 in order to fit  precisely,  not only the AMS-02 positron ratio data, but also 
 the data that are available for the electron and positron flux by AMS-02~\cite{AMS02} and Fermi-LAT~\cite{Fermi_ep} experiments. 
 Having fixed the background we estimate the two main quantities  that generally parametrize  a simplified  DM model:
 the $m_\mathrm{DM}$ and $\vev{\sigma v}$, for the four main  annihilation channels we study, the leptonic 
  $\mu^+ \mu^-$ and  $\tau^+ \tau^-$ and the hadronic  $b \bar{b}$ and $W^+ W^-$. 
   The parameters of any DM annihilation 
fit are determined mainly by the data in the range  $E \gappeq 20$~GeV,
so we use this range in our subsequent fits. One quite important advantage of this choice 
is that in this energy region,  the solar modulation effects are relatively unimportant.

  These issues are discussed in section~\ref{sect:back}.
  In section~\ref{sect:DM} we discuss in details the prediction of these DM models for the  antiprotons and $\gamma$-ray fluxes.
Consequently,  we discuss the effects of these constraints on the various DM models.
Finally, in the  section~\ref{sect:sum} we give a summary of this paper and we discuss the corresponding perspectives.

\section{Background and fits to the electron and positron data}
\label{sect:back}
The first step is to model the background positron fraction induced by cosmic-ray
interactions, using the {\tt GALPROP} version 54~\cite{galprop}. Adjusting its parameters
appropriately, and choosing the solar modulation potential $\Phi = 620$ MeV, a value compatible to the period 
that the experimental data were accumulated.  In any case,  the solar modulation affects mainly the 
low energy region ($E< 1$ GeV), which  as have discussed are not part of our analysis.

 %%%%%%%%%%%%%%%%%galprop points table %%%%%%%%%%%%%%%%%%%%%%%%%%%%%%%%
%\begin{widetext}
\begin{table}[h!]
\begin{center}
\begin{tabular}{l l}
%\hline 
 % {\tt GALPROP} Parameters\\ 
\hline  \hline      
       $D_0$         \; \; \;  & $6.1\times 10^{28}\,  \mathrm{cm^2\, s^{-1}}$ \\ 
       $z_h$           \; \; \;    & 4 kpc     \\ 
       $r_{max}$    \; \; \;   & 25 kpc     \\ 
       $\delta$        \; \; \;  & 0.33     \\        
       $v_{A}$       \; \; \;   & $30 \,  \mathrm { km \, s^{-1}}$    \\
    \hline       \hline        
\end{tabular}
\end{center}
\caption{\it The values for the  {\tt GALPROP} propagation parameters used in our analysis. 
 The diffusion normalization coefficient $D_0$  is given at 4 GeV.}
\label{table:galprop}
\end{table}
%\end{widetext}
%%%%%%%%%%%%%%%%%%%%%%%%%%%%%%%%%%%%%%%%%%%%%%%%%%

The basic choice for the set of the 
{\tt GALPROP} parameters  are  those in~ \cite{Grasso:2009ma}. These   
have been chosen to fit the boron to carbon ratio  B/C. The values of these  
are  similar to the so-called {\tt GALPROP} 2D conventional model~\cite{galprop}, and they are given in Table~\ref{table:galprop}.
In this table, $D_0$ is the diffusion coefficient, $z_h$ and $r_{max}$ are the half-width and maximum size for 2D galactic model respectively,
 $\delta$ is  the index of the power-law dependence of diffusion coefficient and $v_A$ the   Alfv\'{e}n speed.
 The breaking point for the nucleus injection spectral index is chosen to be 9 GeV. The spectral index
above this,  is 2.36 and  below  1.82. For the electrons the  flux normalization at 100 GeV is
$4\times 10^{-7}\, \mathrm{cm^{-2} s^{-1} sr^{-1} GeV^{-1} }$. 
The  breaking point for the electron injection spectrum is located at 4 GeV. Below this point the electron injection spectral index 
is chosen to be $\gamma_0^{el} = 1.6$, like in~\cite{galprop}. 

\begin{figure}[htb]
\begin{center}
\epsfig{file=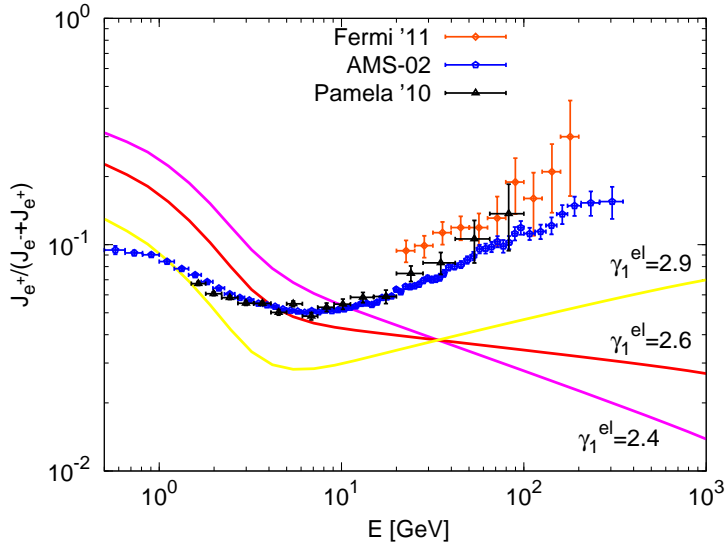,width=0.6\textwidth}

\caption{
\it  The  Fermi-LAT 2011~\cite{Fermi_rat} (orange rectangles),    the AMS-02~\cite{AMS02}  (blue pentagons), 
and the Pamela 2010~\cite{pamela_rat} (black triangles) positron ratio data, 
compared to  various  {\tt GALPROP} background models  for $\gamma_1^\mathrm{el} =2.4$ (purple curve), 2.6 (red curve), and 2.9 (yellow curve).
This range will be used in our analysis. 
}
\label{fig:backg}
\end{center}
\end{figure}

In our analysis, we 
vary the value of the electron injection spectral index above 4 GeV, around the ``conventional" value  $\gamma_1^{el} \simeq  2.5$.
In particular,  we produce eleven background models starting from the value  $\gamma_1^{el} = 2.4 $ up to the 2.9, with step 0.05. 
We are motivated to do so,  because  we notice  in Fig.~\ref{fig:backg} that varying  the  $\gamma_1^{el} $ the slope  of the 
background contribution to the positron ratio changes significantly. 
In particular, larger values of  $\gamma_1^{el} $  provide better alignment to the slope of the AMS-02 data in the high energy region.
These eleven {\tt GALPROP} models will be used as basis to our $\chi^2$ analysis.
 
In Fig.~\ref{fig:backg} we display the three   {\tt GALPROP} models, using  $\gamma_1^\mathrm{el} =2.4$ (purple curve), 
2.6 (red curve)  and 2.9 (yellow curve). This is the range of $\gamma_1^\mathrm{el}$ used in this  analysis. 
In addition we present  the Fermi-LAT 2011~\cite{Fermi_rat} data (orange rectangles), 
 the recent AMS-02~\cite{AMS02} data  (blue pentagons) and the  Pamela 2010~\cite{pamela_rat} (black triangles) positron ratio data.
 Indeed, we noticed  that the large value  $\gamma_1^\mathrm{el}=2.9$ provides better alignment to the AMS-02, 
in the range $E> 10$ GeV. 

The next step is to quantify the dependence of the fit of other data that will be used, on the parameter $\gamma_1^\mathrm{el}$.
In particular,  we estimate the $\chi^2$ including the effect of the DM pair annihilations, 
for  the positron ratio  $J_{e^+}/ (J_{e^-} + J_{e^+}) \equiv J_{e^+}/ (J_{e^\pm})$ as measured  by the AMS-02 experiment~\cite{AMS02}, 
for the electron and positron flux as measured by AMS-02~\cite{AMS02} and Fermi-LAT experiments~\cite{Fermi_ep,Grasso:2009ma} and 
the antiproton ratio $\bar{p}/p$ by the PAMELA~\cite{pamela_p}.  
As discussed earlier, for the fit we use the high energy data ($E\geq20$ GeV).
There are two main reasons for this.   The effects of the DM pair annihilations are expected to be important in this
region. Moreover, this region is not affected significantly from the solar modulation effects.  

Concerning  the DM halo profile,  in our analysis we  study 
two profiles: the  Navarro-Frenk-White (NFW)~\cite{NFW} and the Einasto profile~\cite{einasto}.
The $\chi^2$ analysis and the subsequent results that are related to the electron/positron flux, the positron  and the 
antiproton ratios do  not depend  significantly on the particular choice of the  DM halo profile. Thus, in the 
following the numerical results that are presented are  calculated assuming the Einasto profile. 
On the other hand, we must note that  the choice of the DM profile is important for calculating the $\gamma$-ray flux.
Thus in section~\ref{sect:DM} where we present the constraints that are related to photons we plot 
results both for Einasto and NFW profiles.

\begin{figure}[t!]
\begin{center}
\epsfig{file=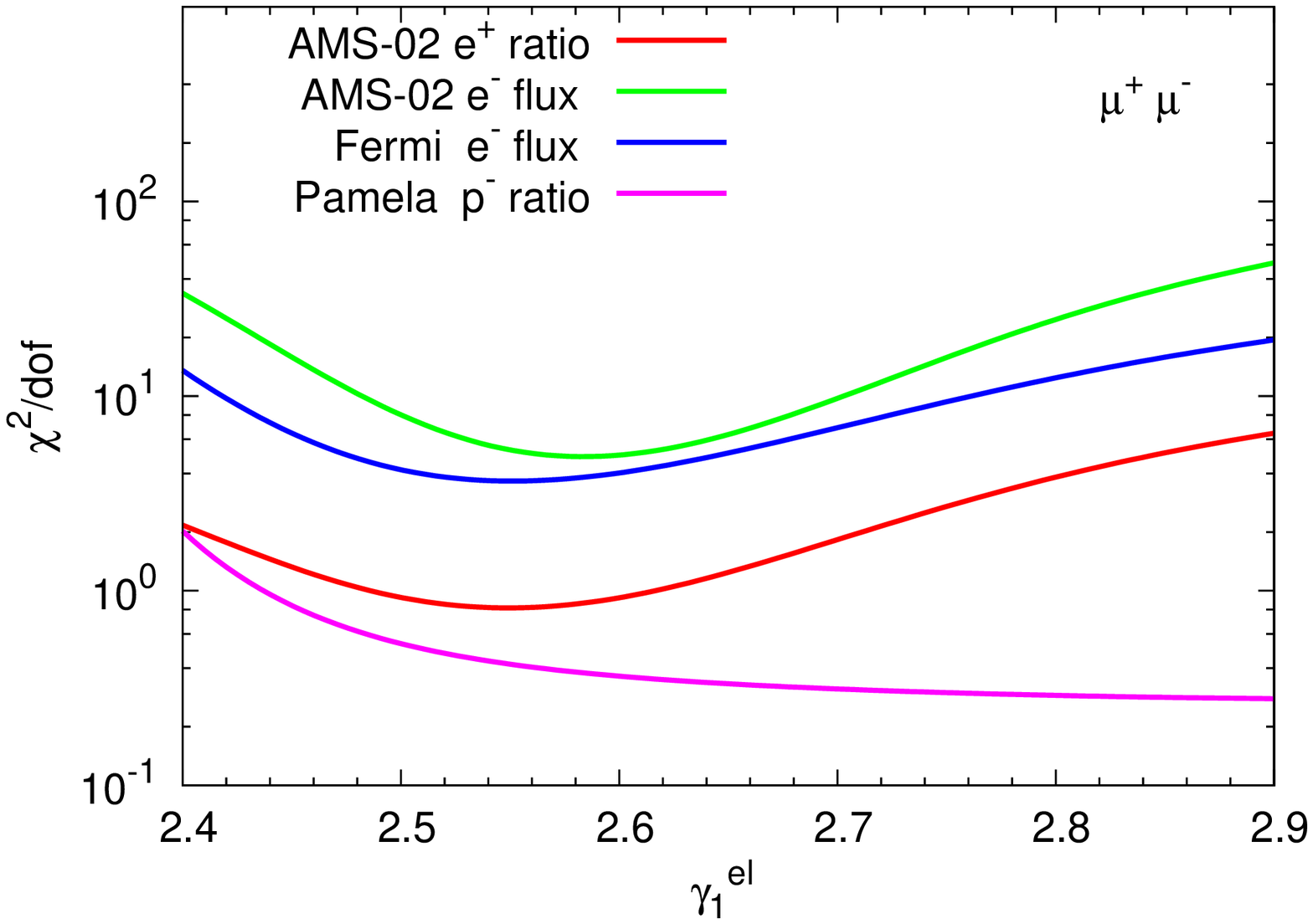,width=0.48\textwidth}
\epsfig{file=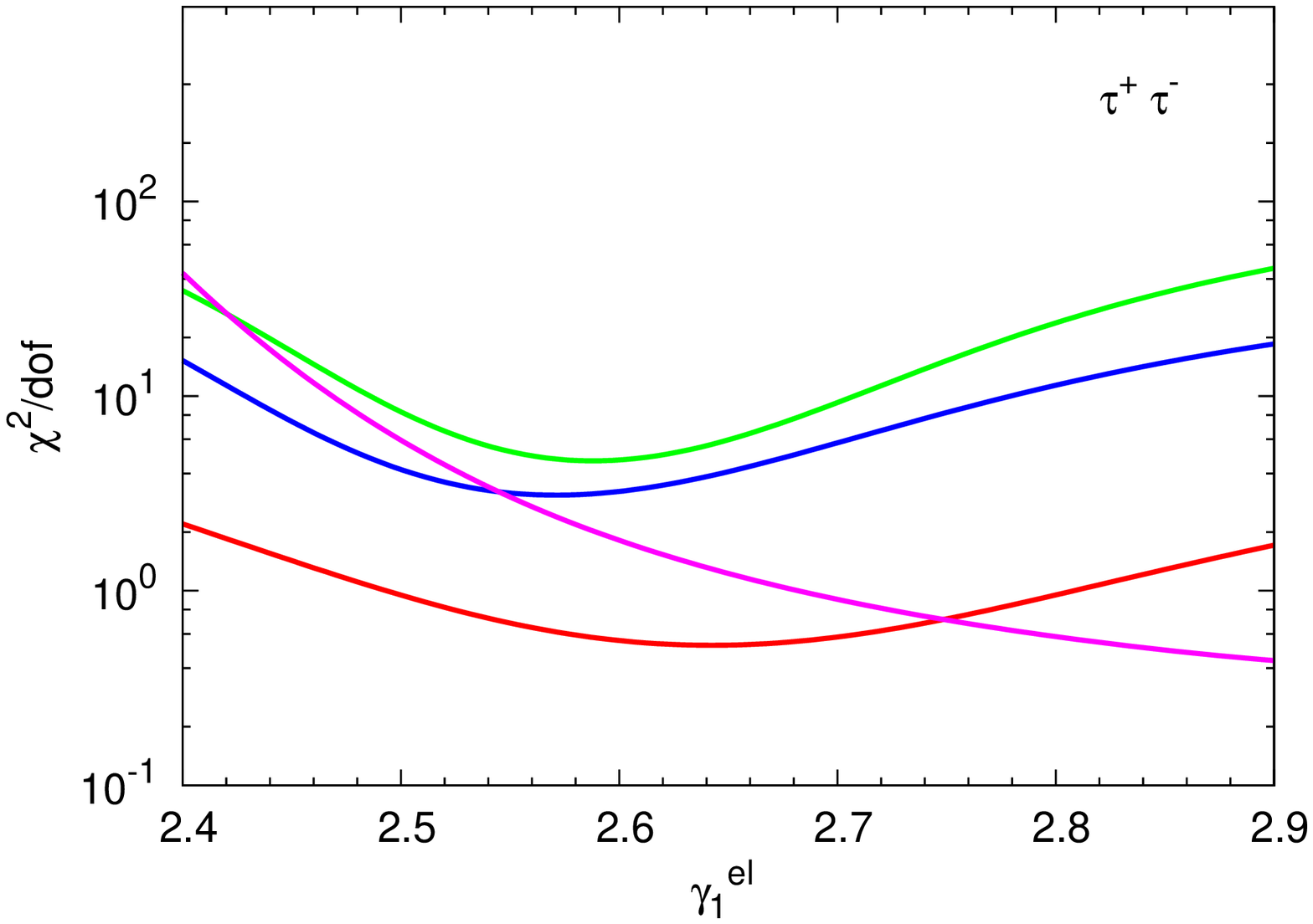,width=0.48\textwidth}\\
\epsfig{file=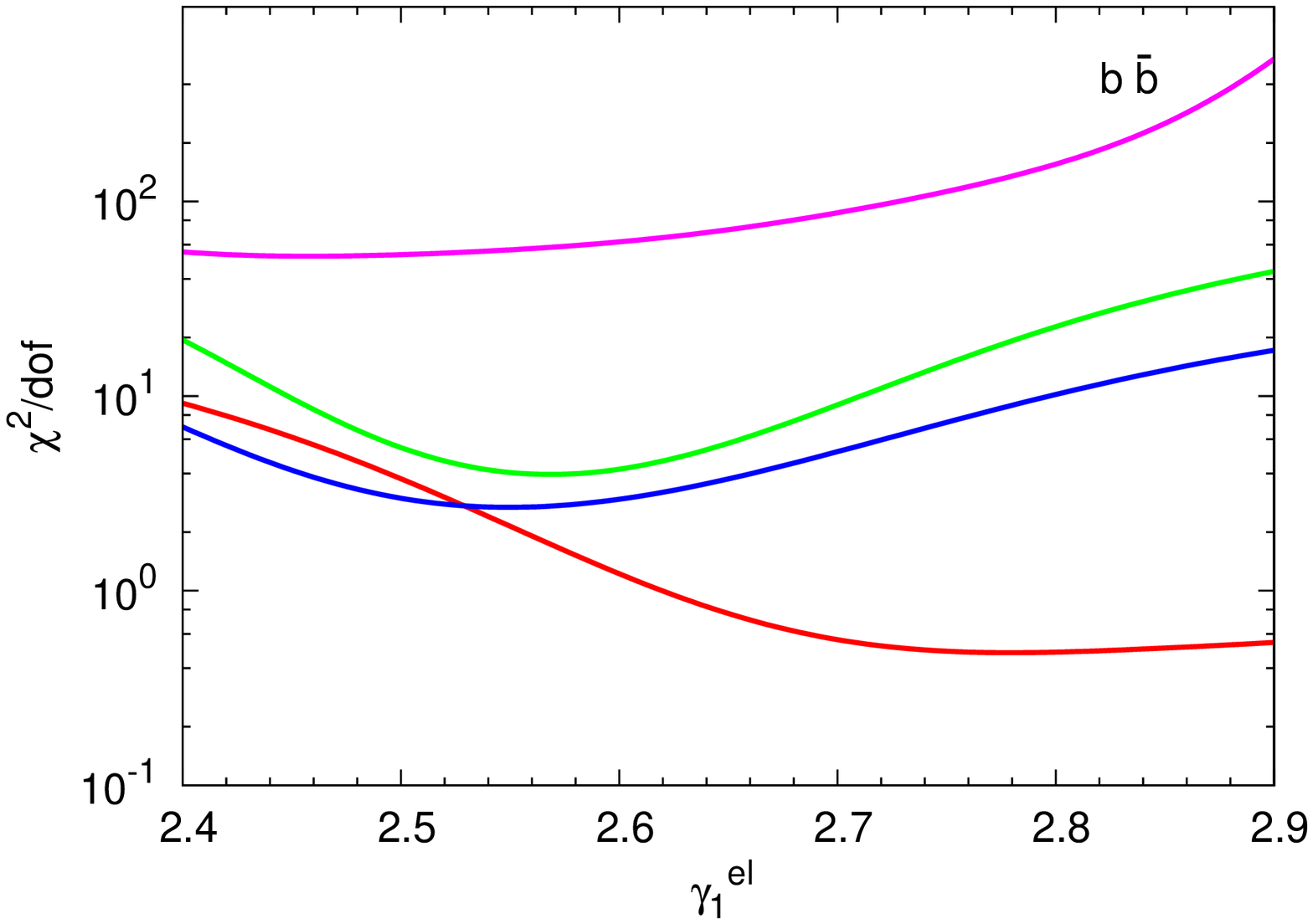,width=0.48\textwidth}
\epsfig{file=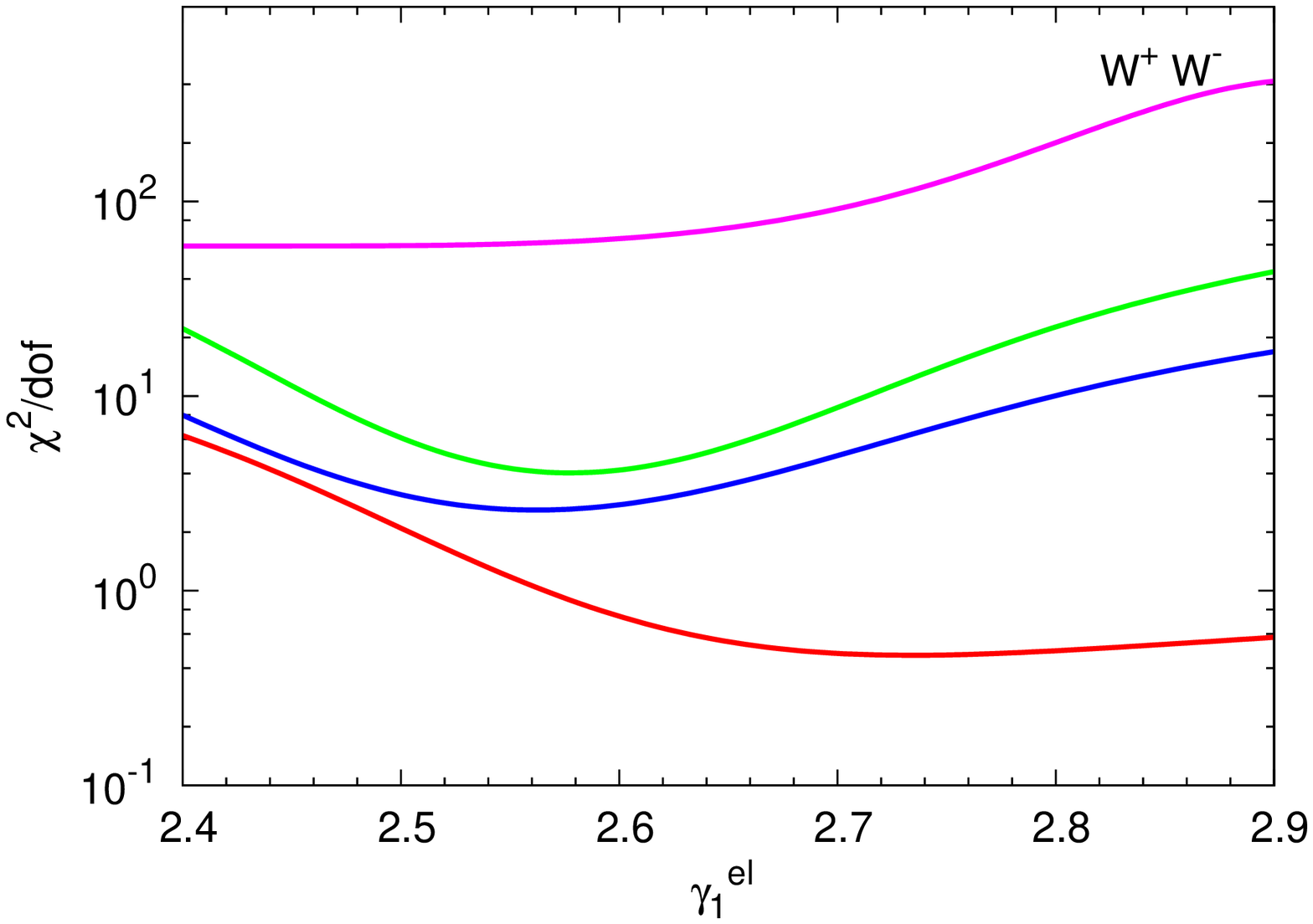,width=0.48\textwidth}
\caption{
\it  The various $\chi^2/dof $ that correspond to the 
 AMS-02 positron ratio data (red curve), to  the electron and positron fluxes data  from AMS-02 (light green curve) and the Fermi-LAT (blue curve), 
and to the antiproton ratio data from PAMELA (purple  curve),  
as a function of the {\tt GALPROP} parameter  $\gamma_1^\mathrm{el} $. The Einasto DM halo profile  is used, the NFW case is
very similar. The  four panels  correspond to the pair annihilation 
channels $\mu^+\mu^-$, $\tau^+\tau^-$,  $W^+ W^-$  and $b \bar{b}$,  clockwise as labeled.
}
\label{fig:g1el}
\end{center}
\end{figure}

Turning now  to our main task,   we vary the $\gamma_1^\mathrm{el} $ and we estimate using {\tt GALPROP} the corresponding 
electron, positron, proton,  antiproton and photon  backgrounds. Using these backgrounds we estimate the $\chi^2$ that
corresponds to the AMS-02 positron ratio data, including the effect of the DM pair annihilations.
As discuss earlier,  the DM
 annihilations  channels we employ are  $\mu^+ \mu^-$,  $\tau^+ \tau^-$, $b \bar{b}$ and $W^+ W^-$.
The {\tt PYTHIA}~\cite{pythia} event simulator is used to model the various spectra of the annihilation  products, that 
will used in our calculation. 
 Since the  $\chi^2$ for the fluxes and ratio we study, depends on the values of $m_\mathrm{DM}$ and $\langle \sigma v \rangle$,
 we would like  to fix for a particular {\tt GALPROP} background the 
  pair ($m_\mathrm{DM}$, $\langle \sigma v \rangle$) that minimize the $\chi^2$ of the AMS-02 positron ratio data.

Using these $m_\mathrm{DM}$ and $\langle \sigma v \rangle$ values 
we plot in Fig.~\ref{fig:g1el}    the $\chi^2$ per degree of freedom (dof, which   is the  energy bin in our case)
as a function of $\gamma_1^\mathrm{el} $ for the four DM annihilation channels.
We calculate the $\chi^2/dof$ for the AMS-02 positron ratio data~\cite{AMS02}, 
for  the electron and positron fluxes data  from AMS-02~\cite{AMS02} 
and    Fermi-LAT~\cite{Fermi_ep},  and finally 
for the antiproton ratio data from PAMELA~\cite{pamela_p}. 
Consequently, one can use the curves at  Fig.~\ref{fig:g1el} to fix the $\gamma_1^\mathrm{el} $ parameter. 
We notice  indeed that, as the $\gamma_1^\mathrm{el} $ varies,
 there is a region around $\gamma_1^\mathrm{el} \simeq 2.6 $ where the  fit for the electron/positron flux  data and  
  for the AMS-02 positron ratio data is quite satisfactory.  
  
  In this region the $\chi^2$ is less than  $10$ for the AMS-02 $e^-$ data (light green curve) and slightly 
  better for the Fermi-LAT (blue curve). Moreover, the  $\chi^2/dof$ is below unity for the AMS-02 positron ration (red curve).
   This happens    for all the DM channels we study. 
 We notice that   the better  $\chi^2$ for the AMS-02 positron ratio data is not an accident, since  the
   chosen values for  $m_\mathrm{DM}$ and $\langle \sigma v \rangle$ are those that minimize it.
  On the other hand, the quality of the antiproton ratio data fit doesn't depend on the $\gamma_1^\mathrm{el}$  parameter and is satisfactory 
   for the leptonic channels ($\mu^+ \mu^-$, $\tau^+ \tau^-$), everywhere. 
 As it was expected, on the other hand, is  quite bad for the hadronic channels ($b \bar{b}$, $W^+ W^-$).
Therefore, the safest choice for the {\tt GALPROP } background is to fix   $\gamma_1^\mathrm{el} = 2.6 $
and this will be done in the following.

%\clearpage
\section{DM interpretation of the AMS-02 positron data }
\label{sect:DM}
%%\subsection{$\tau^+ \, \tau^-$ channel}

Having fixed the {\tt GALPROP} background by choosing $\gamma_1^\mathrm{el} = 2.6 $, 
one can evaluate the corresponding values of $m_\mathrm{DM}$ and $\langle \sigma v \rangle$ that minimize 
the $\chi^2$ for the AMS-02 positron ratio data, as well as the electron and positron flux data from the same experiment. 
These values for the four  DM annihilation channels are given in Table~\ref{table:bestfit}, 
and these will used in the our analysis. From this Table is evident that the two leptonic channels 
favor smaller values for $(m_\mathrm{DM},\langle \sigma v \rangle)$, while the two hadronic 
channels much larger.

 %%%%%%%%%%%%%%%%% sig mdm  fit %%%%%%%%%%%%%%%%%%%%%%%%%%%%%%%%
%\begin{widetext}
\begin{table}[h!]
\begin{center}
\begin{tabular}{   l l r  }
 \hline   \hline   \vphantom{\rule{0pt}{12pt}}
       channel                  & \;   \; \;   \;   $\langle \sigma v \rangle \,  [\mathrm{cm^3/s}] $&\;   \;   \;  \; $m_\mathrm{DM} \, [\mathrm{GeV}]$  
        \vphantom{\rule{0pt}{12pt}}  \\ \hline   \vphantom{\rule{0pt}{12pt}} 
       $\mu^+ \mu^-$      & \;    \; \;   \;  $7.50 \times 10^{-24}$ & 465 \;     \\ 
       $\tau^+ \tau^-$     &  \;  \; \;   \;  $8.91 \times 10^{-23}$  &\ 1758 \;      \\ 
       $b \, \bar{b}$           &  \;  \; \;   \;   $2.99 \times 10^{-21}$ & 58546  \;     \\        
       $W^+ W^-$          & \;   \; \;   \;  $1.12 \times 10^{-20}$  & 91728  \;          \\ 
     \hline       \hline         
\end{tabular}
\end{center}
\caption{\it The values of  $\langle \sigma v \rangle$ and  $m_\mathrm{DM}$ for the pair annihilation channels
we discuss, for  $\gamma_1^\mathrm{el} =2.6$.}
\label{table:bestfit}
\end{table}
%\end{widetext}
%%%%%%%%%%%%%%%%%%%%%%%%%%%%%%%%%%%%%%%%%%%%%%%%%%

\begin{figure}[t!]
\begin{center}
\epsfig{file=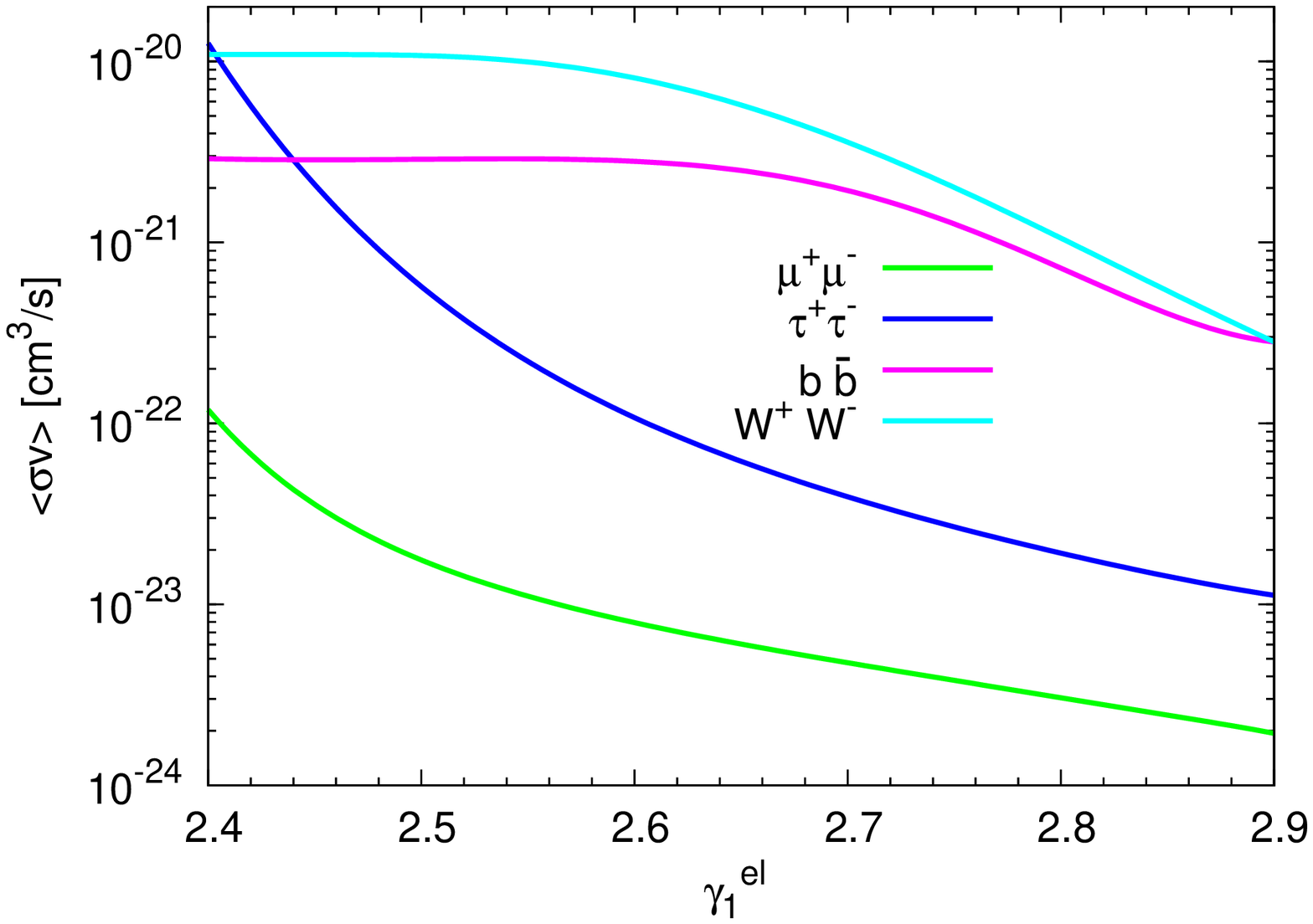,width=0.49\textwidth}
\epsfig{file=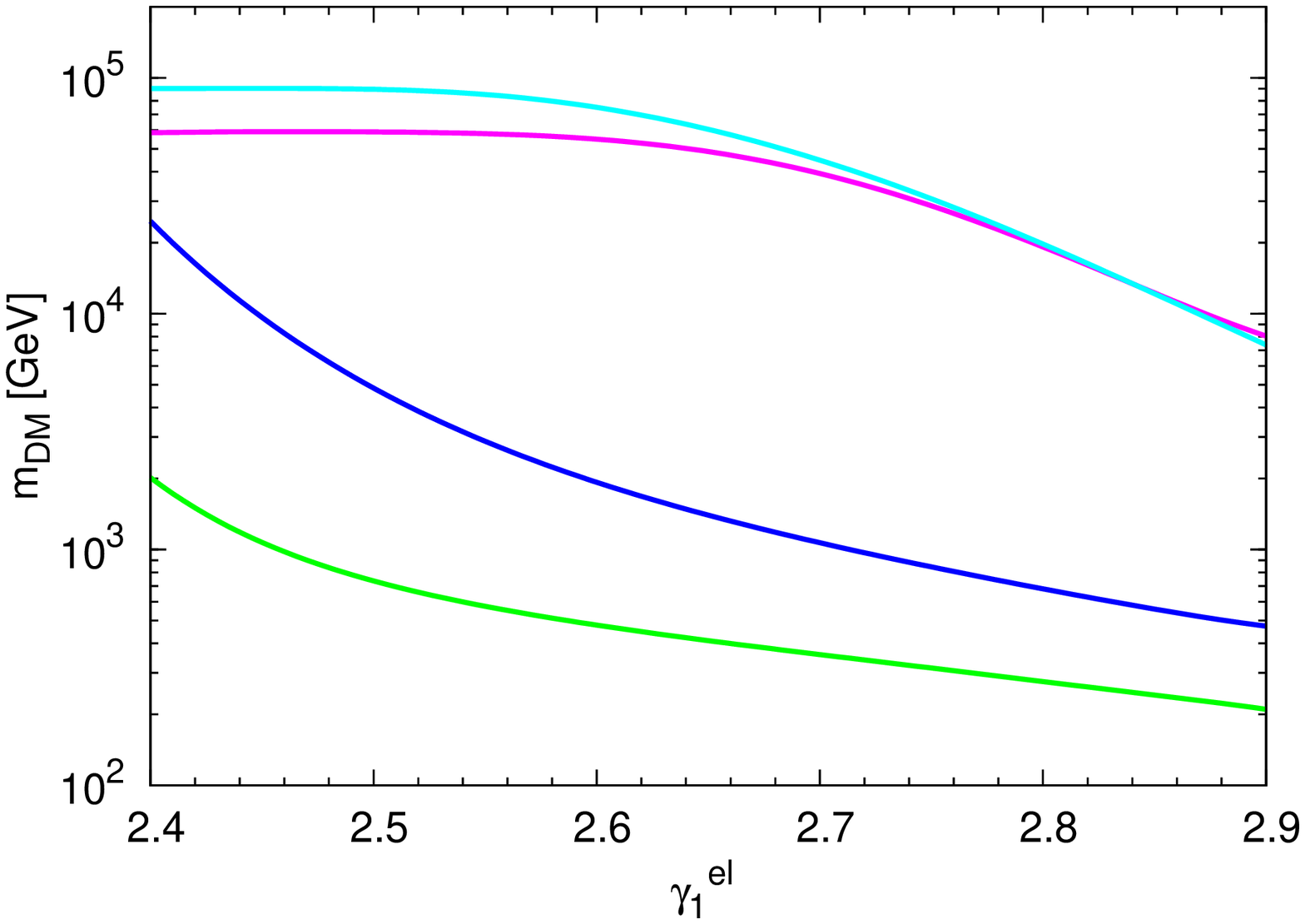,width=0.49\textwidth}
\caption{
\it  The values of $\langle \sigma v \rangle$ (left plot) and  $m_\mathrm{DM}$ (right plot)  that minimize the $\chi^2$ of  the positron 
ratio data from  AMS-02, as a function of    {\tt GALPROP}  parameter $\gamma_1^\mathrm{el} $. We plot  these  for the
four decay channels under study:  $\mu^+\mu^-$ (light green curve), $\tau^+\tau^-$ (blue curve), $b \bar{b}$ (purple curve) and $W^+ W^-$  (cyan  curve).
For both plots the Einasto DM halo profile  is assumed, and again the NFW case is
quite  similar. 
}
\label{fig:fit_mdmsig}
\end{center}
\end{figure}

For completeness we display in Fig.~\ref{fig:fit_mdmsig} the dependence of the   $\langle \sigma v \rangle$ (left figure)
and $m_\mathrm{DM}$ (right figure) on  $\gamma_1^\mathrm{el} $, for the annihilation channels we study. 
In particular, the values displayed in Table~\ref{table:bestfit} correspond to the preferred   value $\gamma_1^\mathrm{el} = 2.6 $.
Using these values we have prepared various figures that illustrate  the 
quality of the fits to the  electron, positron and antiproton data.

\begin{figure}[htb]
\begin{center}
\epsfig{file=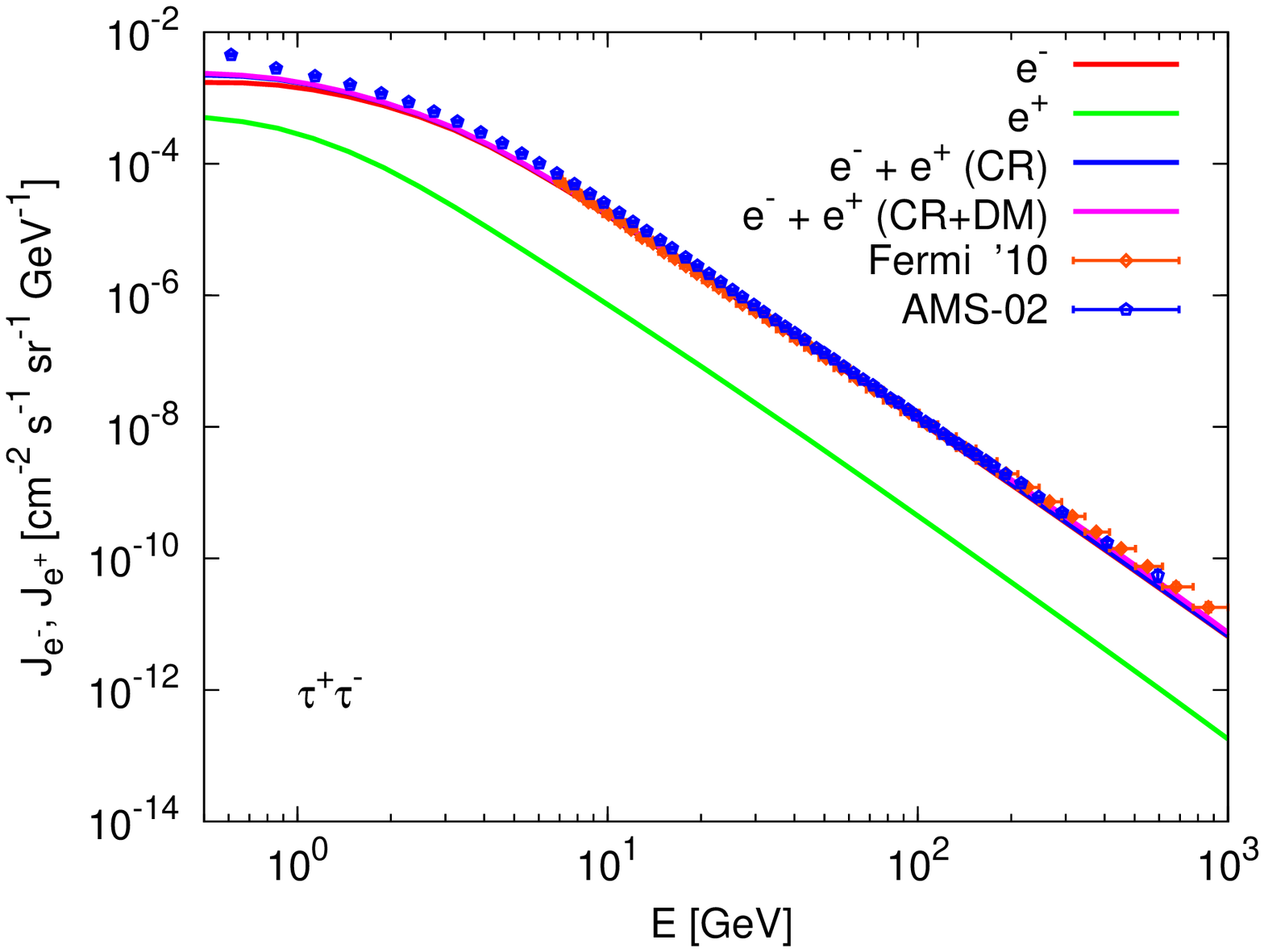,width=0.496\textwidth}
\epsfig{file=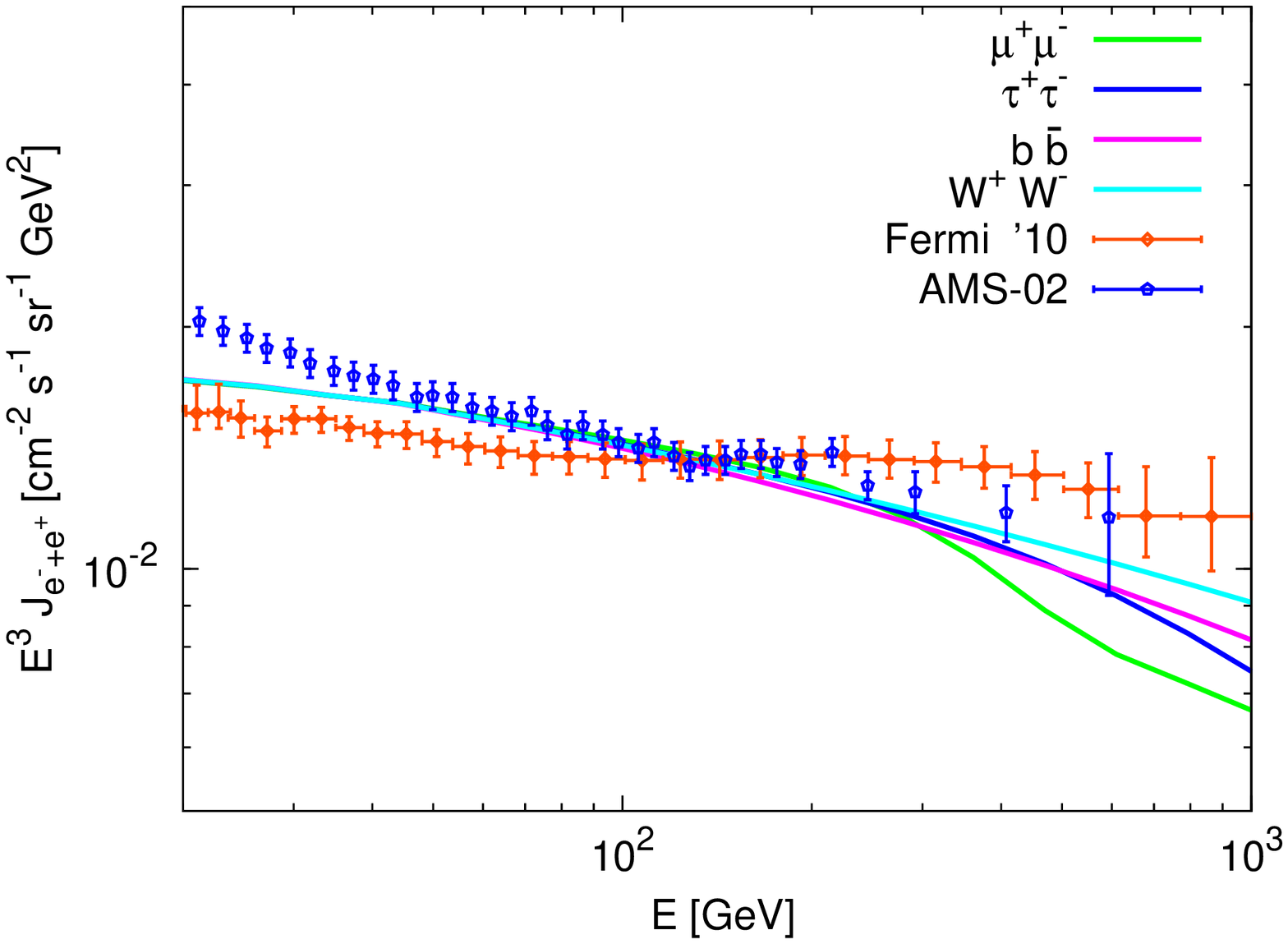,width=0.496\textwidth}
\caption{
\it  Left panel:   The red curve is the electron  $J_{e^-}$ flux, the light green curve is the positron   $J_{e^+}$ flux and the blue curve is the 
  electron/positron  $J_{e^+} + J_{e^-} =   J_{e^\pm}$  flux from the CR background only,
  assuming  $\gamma_1^\mathrm{el} = 2.6 $.    The purple line is the $J_{e^\pm}$ from the CR and the DM contribution (using 
  the values of Table~\ref{table:bestfit}). 
  We have also plotted the corresponding $J_{e^\pm}$  data from the Fermi-LAT (orange rectangles) and AMS-02 (blue pentagons) experiments. 
  In this figure we assume the $\tau^+ \tau^-$ annihilation channel. 
  Right panel: We focus at the energy region $20 \GeV -  1 \TeV$ and we display the quantity $E^3 J_{e^\pm}$ for the individual channels:
 $\mu^+\mu^-$ (light green), $\tau^+\tau^-$ (blue),  $b \bar{b}$ (purple)  and  $W^+ W^-$ (cyan) for the same background/DM point as in the
 left plot.  For both figures  we assume the  Einasto halo profile. 
}
\label{fig:fit_elpo}
\end{center}
\end{figure}

In Fig.~\ref{fig:fit_elpo} (left panel) we display  the electron  $J_{e^-}$ (red curve),  the positron   $J_{e^+}$  (light green curve) and the 
total $J_{e^+} + J_{e^-} =   J_{e^\pm}$ (blue curve)  flux  from the CR background only, using $\gamma_1^\mathrm{el} = 2.6 $,
as discussed before. In addition,  the   purple line incorporates the effects for the  DM annihilations assuming in particular 
the $\tau^+ \tau^-$ channel, using the values of Table~\ref{table:bestfit}). 
One can see that in the range $E \geq 20 $ GeV,  used   in our analysis,  
  the total flux describes quite well  the data from AMS-02\cite{AMS02} and Fermi-LAT\cite{Fermi_ep}. 
 For the sake of clarity,  in the right panel 
 we  focused in the energy range $E > 20$ GeV  
 and we display the quantity $E^3 J_{e^\pm}$ for all the four DM annihilation channels.
In the region $E<100$ GeV,  where there is an evident disagreement  between the AMS-02 and the Fermi-LAT data,
the total fluxes  are placed  between these data. On the other hand,  in the
region $E>100$ GeV the  fluxes appear to fit  better the AMS-02 data.

\begin{figure}[htb]
\begin{center}
\epsfig{file=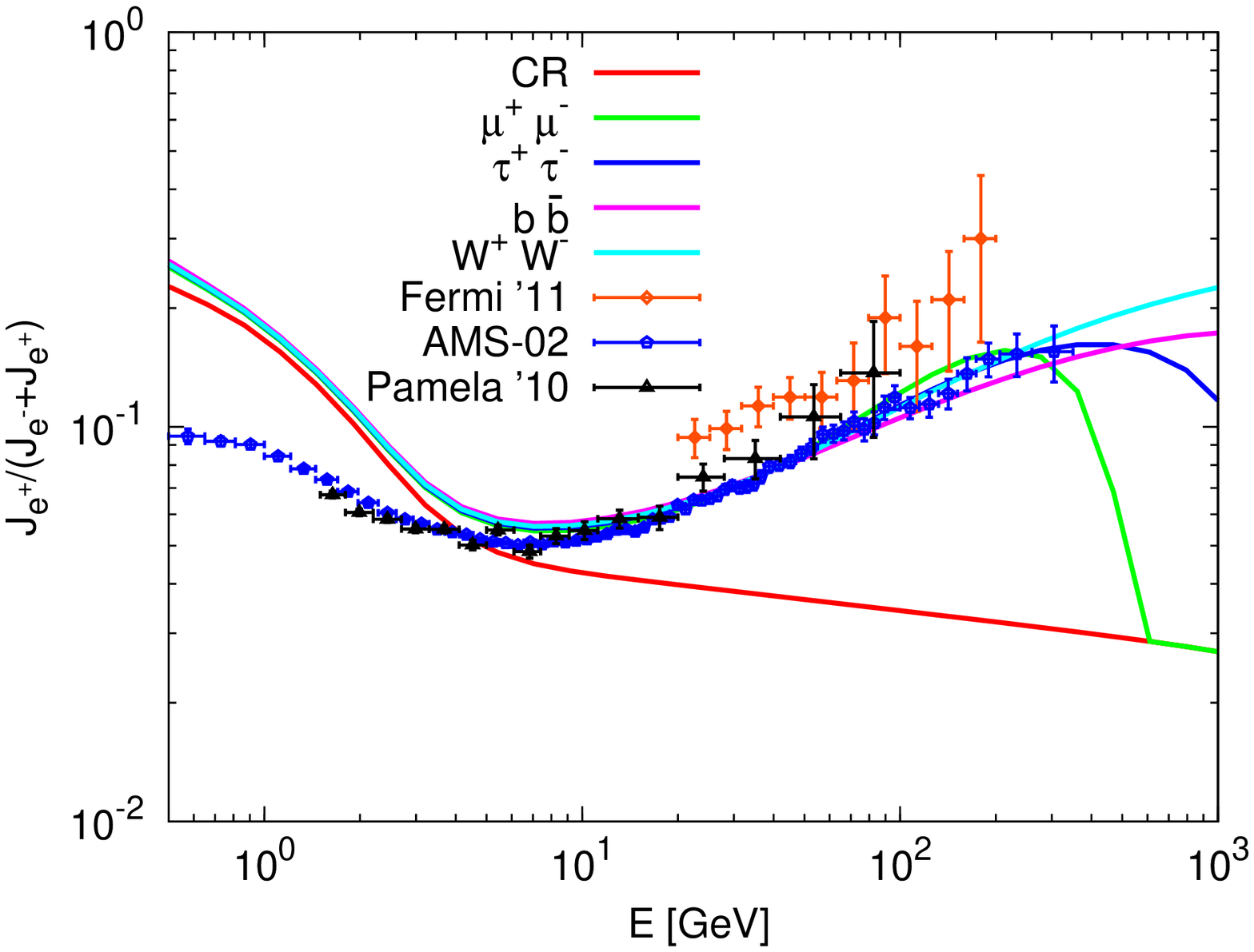, width=0.496\textwidth} 
\epsfig{file=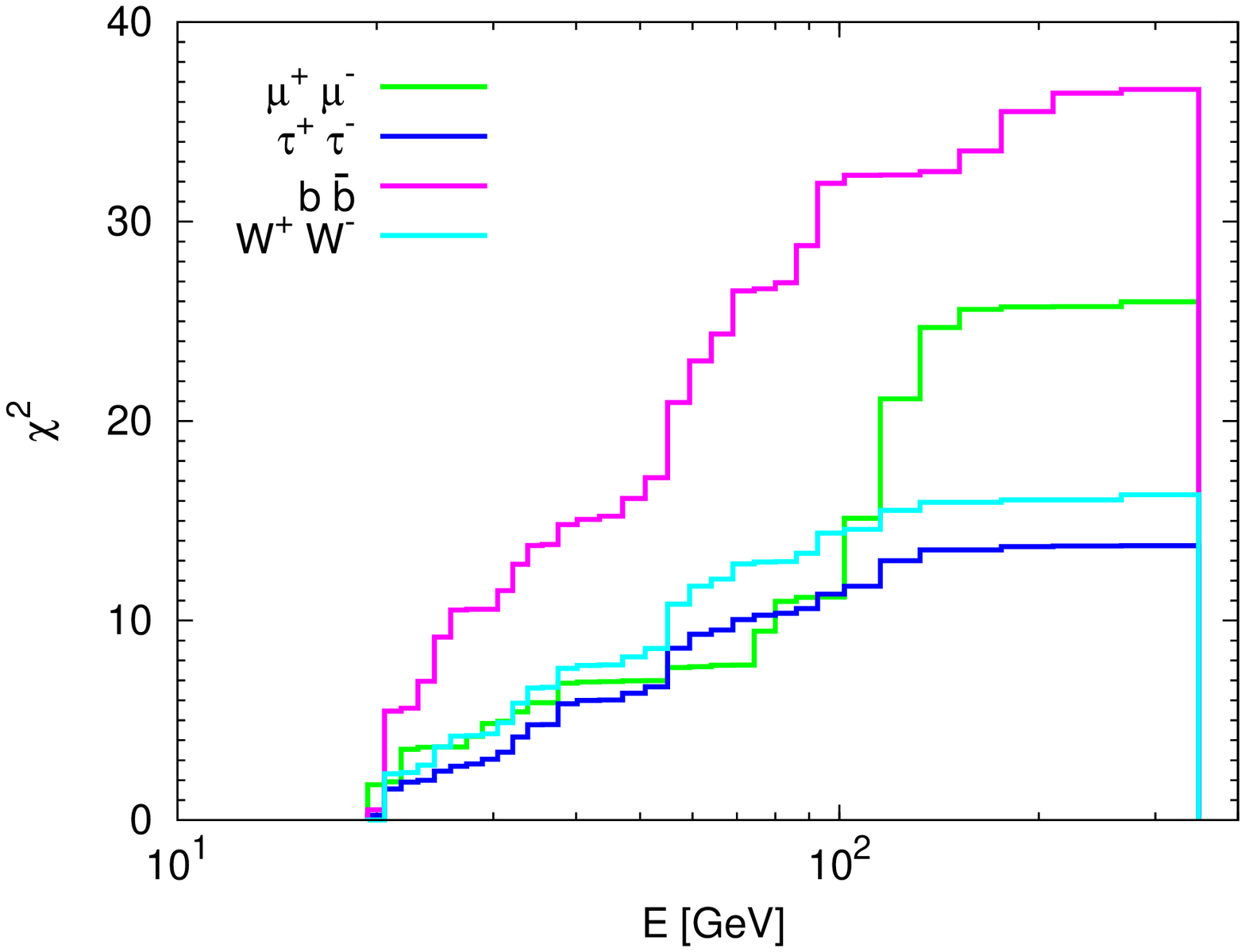,width=0.496\textwidth}
\caption{
\it  Left panel: Positron ratio against the AMS-02 data: The {\tt GALPROP} CR background  using    $\gamma_1^\mathrm{el} = 2.6 $ as described in the text 
(red curve).   
Including the DM channel $\mu^+\mu^-$  (light green), $\tau^+\tau^-$ (blue),  $b \bar{b}$ (purple)  and  $W^+ W^-$ (cyan) for the same background,
using the values of  of Table~\ref{table:bestfit}. 
Like in Fig.~\ref{fig:backg} we plot in addition
 the corresponding positron ratio   data from the Fermi-LAT (orange rectangles), the AMS-02 (blue pentagons) and 
Pamela (black triangles) experiments. 
Right panel: The $\chi^2$  that corresponds to the  curves of the  left panel for the pair annihilation channels  
$\mu^+\mu^-$  (light green), $\tau^+\tau^-$ (blue),  $b \bar{b}$ (purple)  and  $W^+ W^-$ (cyan), for $E \geq 20 \GeV$. 
 We assume the  Einasto halo profile. 
}
\label{fig:ratio_ein}
\end{center}
\end{figure}

In Fig.~\ref{fig:ratio_ein} we present the fits for the  positron ratio data.   As  expected, in the region $E \geq 20$ GeV
where  our fit is  performed, the predicted ratio describes  the AMS-02 data quite well.
In the region $E=300-400$ GeV we notice that the positron ratio for  leptonic channels $\mu^+\mu^-$ and $\tau^+\tau^-$ has a maximum. 
In particular,   for the  tau case is maximum  is wider. On the other hand, the two hadronic channels  ($b\bar{b}$ and $W^+ W^-$)
predict positron ratio that  rise continuously  in this region. 
The corresponding  $\chi^2$'s  for the fit to AMS-02 positron ratio data are displayed in the left panel in Fig.~\ref{fig:ratio_ein}.
Taking into  account that for the  AMS-02 data  32 energy bins are used in the region $E > 20$ GeV,  
the $\chi^2/dof$ appears to be of order of one or less. The $b\bar{b}$ channel appears to have slightly bigger $\chi^2$, while 
the $\tau^+\tau^-$ data fits much better the data.

\begin{figure}[h!]
\begin{center}
\epsfig{file=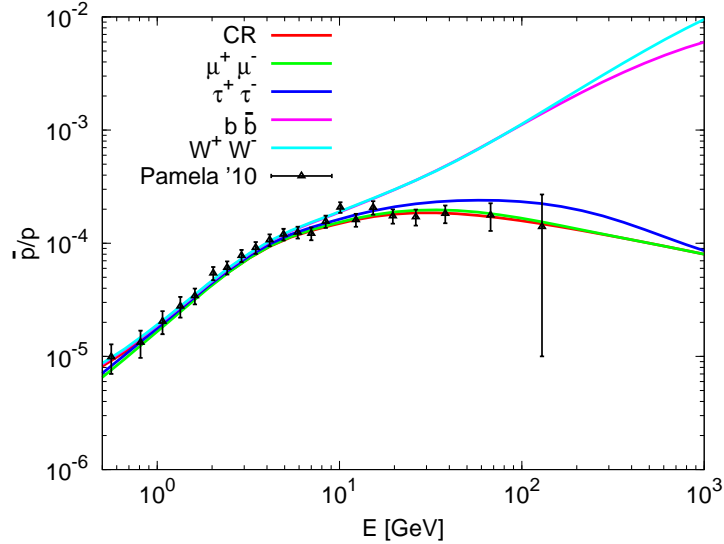, width=0.6\textwidth}
\caption{
\it  The $\bar{p}/p$ ratio assuming  the values of $m_\mathrm{DM}$ and  $\langle \sigma v \rangle$ and
the background  that minimize the $\chi^2$, using the values of  of Table~\ref{table:bestfit},  as in Fig.~\ref{fig:ratio_ein}.
We have also plotted the corresponding  Pamela data (black triangles). 
 We assume the  Einasto halo profile.  
}
\label{fig:antip}
\end{center}
\end{figure}

The $\bar{p}/p$ fraction  for the DM annihilation channels we study are displayed   in the Fig.~\ref{fig:antip}. 
The leptonic channels doesn't produce hadrons directly, but they do 
produce protons and antiprotons 
through the W/Z-strahlung associated  processes, as described  in~\cite{WZ_brems}. 
The produced antiprotons for those channels
can fit quite well the experimental data from PAMELA~\cite{pamela_p}. On the other hand,
as  discussed already,  the hadronic channels  ($b\bar{b}$ and $W^+ W^-$) produce to many 
antiprotons and they are compatible to  the data.

\begin{figure}[t!]
\begin{center}
\epsfig{file=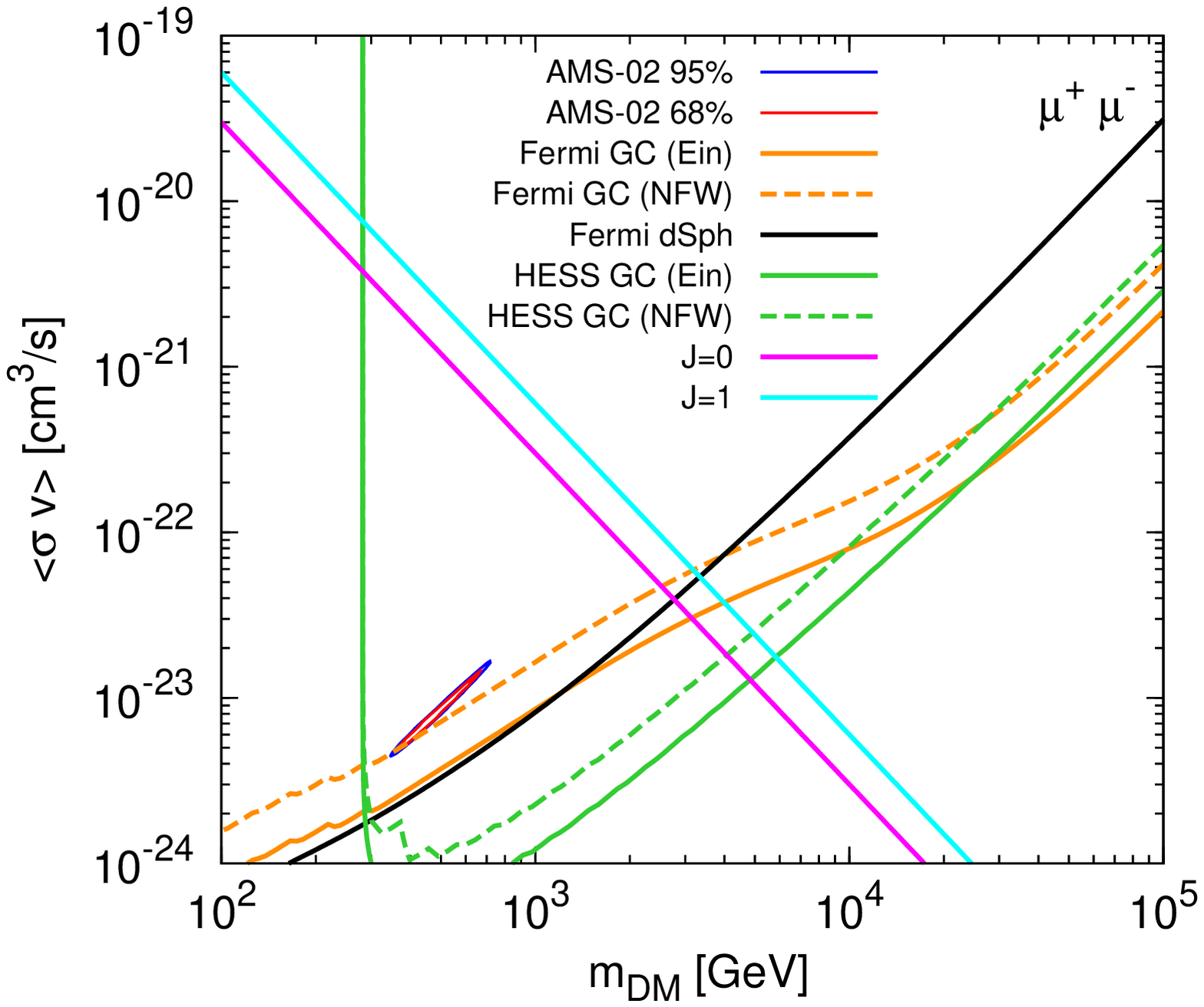,width=0.488\textwidth}
\epsfig{file=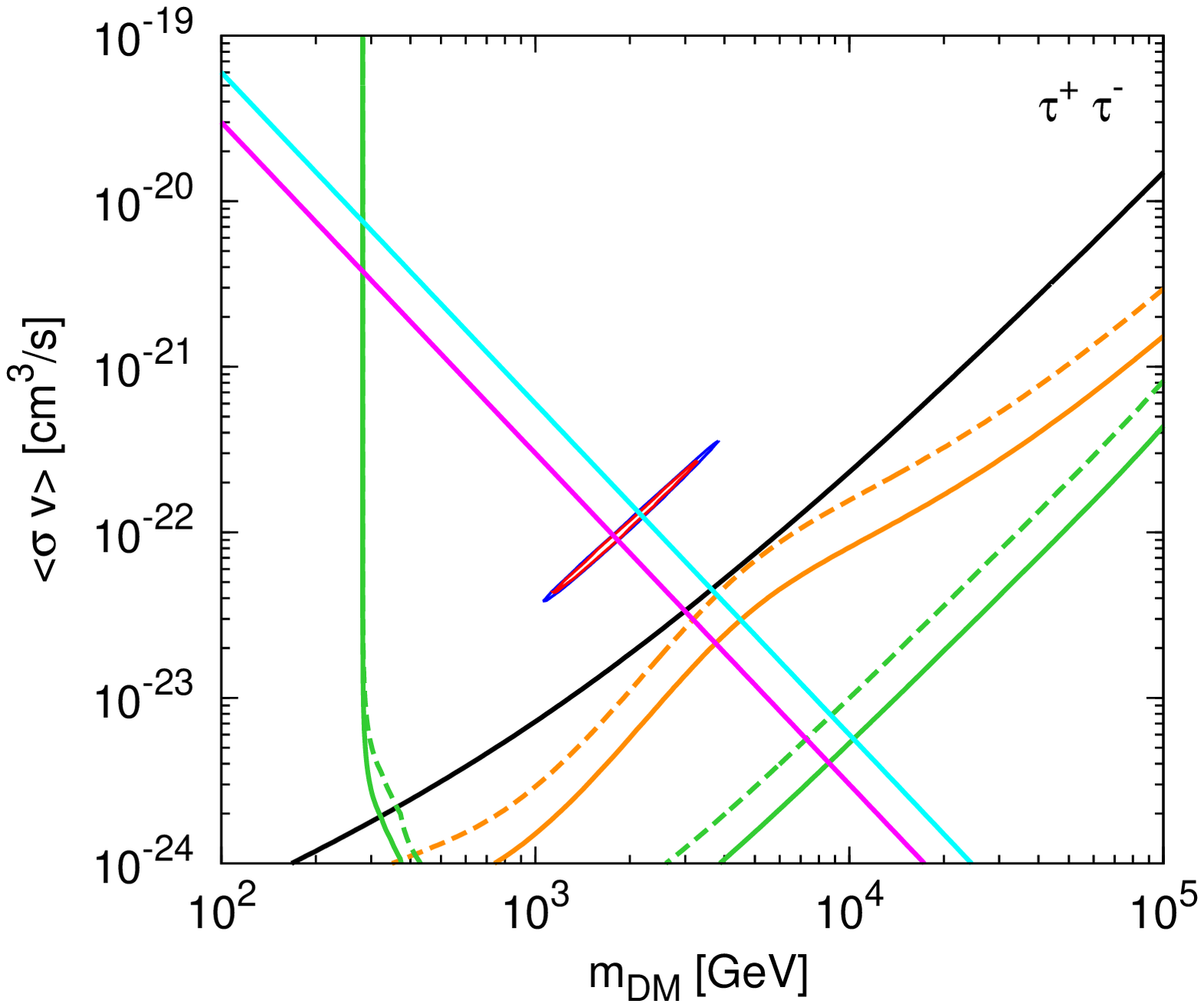,width=0.488\textwidth} \\
\epsfig{file=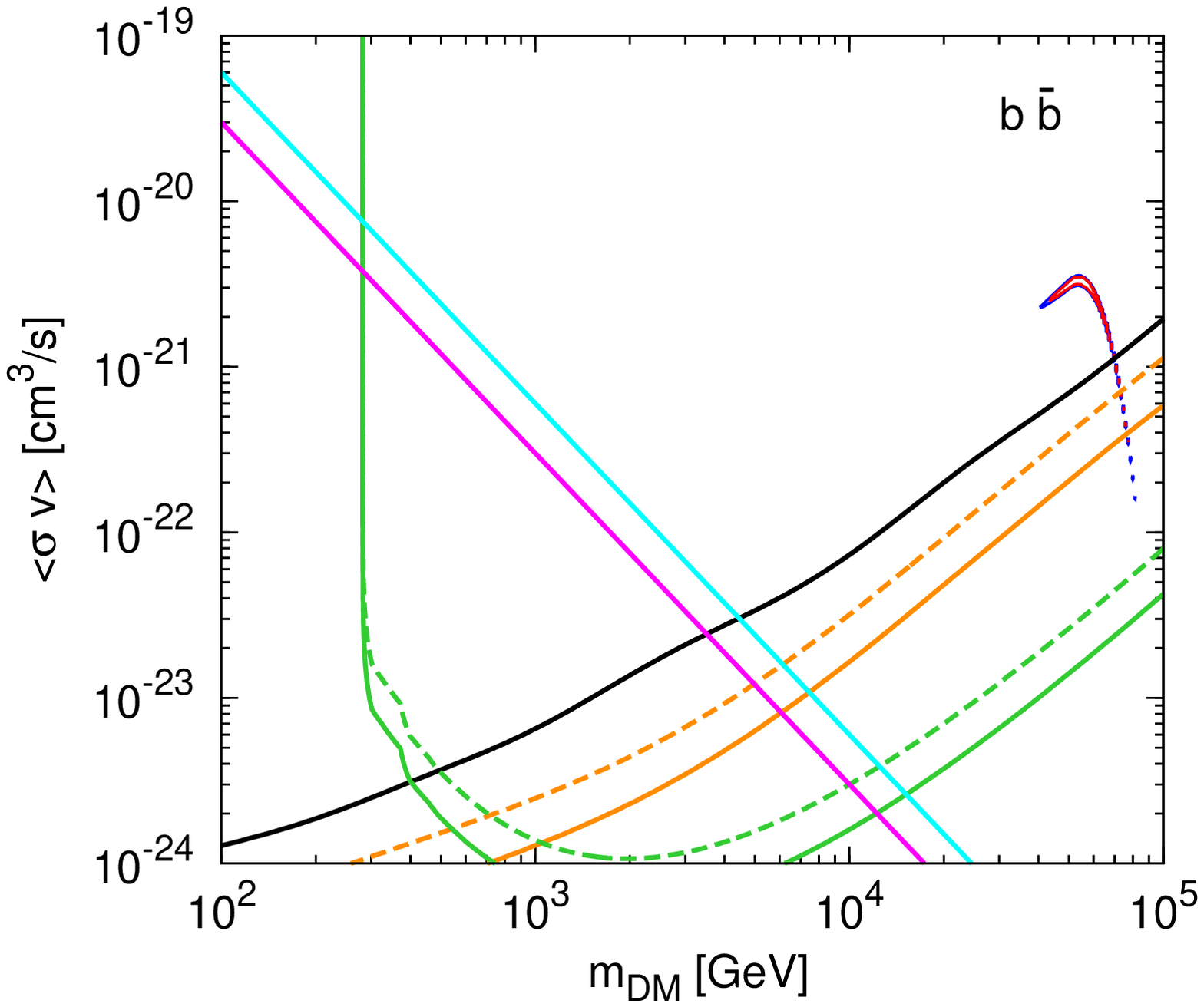,width=0.488\textwidth}
\epsfig{file=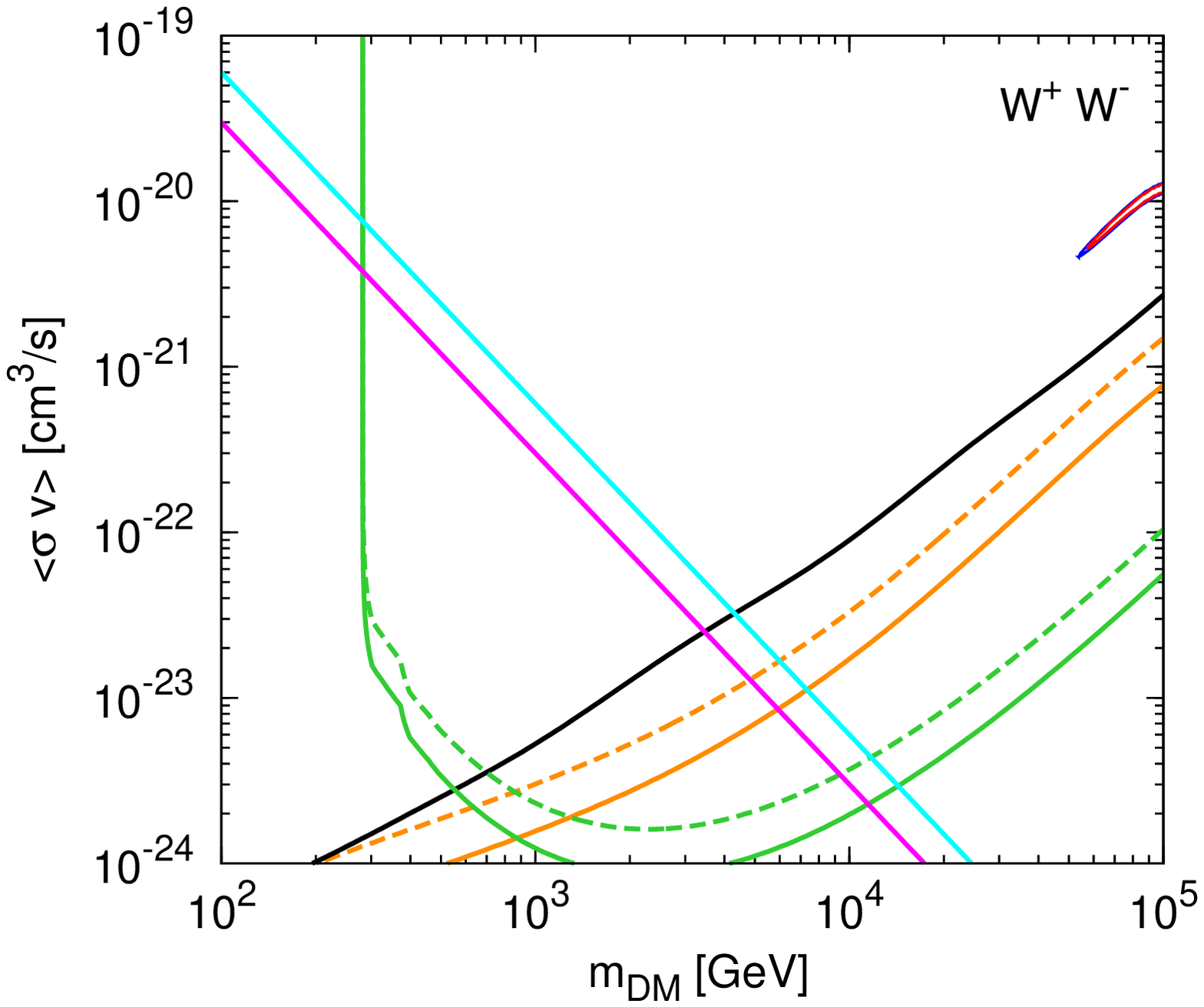,width=0.488\textwidth}
\caption{
\it  The 68\% (red curve) and 95\% (blue curve) confidence level contours in the { \rm{($m_\mathrm{DM}$}, $\langle \sigma v \rangle$)} plane that 
correspond to the AMS-02 positron ratio data. 
The orange solid (dashed) curve corresponds to the Fermi-LAT data from the galactic center  assuming 
the Einasto  (NFW) profie.
The  black curve marks the exclusion limit  using the Fermi-LAT data corresponding to the  dwarfs galaxies.
 The solid (dashed)  dark  green curve marks   the  H.E.S.S  galactic center exclusion limit  using again  
the Einasto  (NFW) profie.
Finally,  the purple (cyan) curve corresponds to the unitarity $J=0$ ($J=1$) limit, as described in the text.
}
\label{fig:ct}
\end{center}
\end{figure}

In Fig.~\ref{fig:ct} we display the  68\% (red curve) and 95\% (blue curve) confidence level (CL) contours in the { \rm{($m_\mathrm{DM}$}, $\langle \sigma v \rangle$)} plane, for the four DM annihilation channels under study:  $\mu^+\mu^-$, $\tau^+\tau^-$,  $W^+ W^-$  and $b \bar{b}$,  clockwise as labeled.
As before,  we use the EInasto halo profile 
to delineate these 
 CL contours, and we the corresponding contours for the NFW profile is very similar. 
The central points of these regions are the points from the  Table~\ref{table:bestfit}.
In addition,   we have marked using a purple (cyan) line the unitarity limit assuming  $J=0$ ($J=1$).
Moreover we have plotted various curves that mark the region that excluded due 
to $\gamma$-ray data from the Fermi-LAT and H.E.S.S~\cite{HESS,HESS2} experiments. 
In particular, the orange solid (dashed) curve corresponds to the Fermi-LAT data from the galactic center (GC) following
the method of~\cite{EOSgamma},  assuming 
the Einasto  (NFW) profie.
The  black curve marks the exclusion limit  using the Fermi-LAT data corresponding to dSph~\cite{dsphfermi}.
 The solid (dashed)  dark  green curve marks   the  H.E.S.S~\cite{HESS} galactic center exclusion limit  using again  
the Einasto  (NFW) profie.
The Fermi-LAT experiment can detect $\gamma$-rays in the energy range 20 MeV to 300 GeV.  
On the other hand H.E.S.S detector  is sensitive to more energetic photons, from few hundreds GeVs  up to few tens of TeVs.
In principle, the H.E.S.S limit  is more relevant to the our analysis, since AMS-02 positron ratio data
favour rather large values of $m_\mathrm{DM}$ and $\langle \sigma v \rangle$.

For this study  we use the  
Fermi-LAT data recorded since the beginning  of the mission  (August 4th, 2008)  until very recently  (Dec 10th, 2013). 
We use the package  ScienceTools~\cite{tools} as it is suggested by the collaboration in order to analyze the Fermi-LAT data.
Especially for the dSph constraints 
we combine  data from ten dSph galaxies: Bootes I, Carina, Coma Berenices, Draco, Fornax, Sculptor, Segue 1, Sextans, Ursa Major II and Ursa Minor, 
as it was done in  ~\cite{dsphfermi}, and using the $J$-factors for dSph DM halos used in this analysis, we calculate the combined  95\% CL upper limits shown in Fig.~\ref{fig:ct},  using  the joint likelihood of these dSph.  The excluded region is in agreement with Fig. 2 presented in   ~\cite{dsphfermi}.  
To construct  the  Fermi-LAT bound related to the GC we focus to  a  7-degree window centered  
at the position of the brightest source in the GC: 
$\mathrm{RA}=266.46^0$, $\mathrm{Dec}=-28.97^0$, as in~\cite{Vitale:2009hr}.

As it is apparent from Fig.~\ref{fig:ct} the two leptonic decay channels: $\mu^+\mu^-$ and the central region of the   $\tau^+\tau^-$ do not violate 
the unitarity bounds. On the other hand, the two hadronic channels:  $W^+ W^-$  and $b \bar{b}$ do violate them.  
Concerning the $\gamma$-ray bounds from Fermi-LAT (GC and dSph) and H.E.S.S (GC) the $\mu^+\mu^-$,  $\tau^+\tau^-$ and $W^+ W^-$  are excluded at the  95\% CL using these bounds  assuming either the NFW or the Einasto halo profile. Only a small tail from the $b \bar{b}$ AMS-02
region can be compatible to the Fermi-LAT bounds from GC or dSph, but again is excluded due to H.E.S.S GC data.

\newpage
\section{Summary and Prospects}
\label{sect:sum}
We have explored  the possibility to explain the recent AMS-02 positron ratio data
using the effects that related to the DM annihilations. We study two leptonic channels 
$\mu^+ \mu^-$, $\tau^+ \tau^-$  and two hadronic $b \bar{b}$, $W^+ W^-$. We employ 
for our analysis both Einasto and NFW halo DM profiles.

 We calculate the CR background for electron, positron, protons, antiprotons and photons using the 
 {\tt GALPROP} code. We perform a $\chi^2$ analysis based on the 
 the {\tt GALPROP} conventional model, that is known to be compatible to B/C data.
 We fix the CR background parameters in such way that minimize the $\chi^2$ not
 only for the AMS-02 positron ratio data, but also for
 the data that are available for the electron and positron flux from AMS-02 and Fermi-LAT collaborations. 
 As basis for our $\chi^2$ analysis we use the high energy bins with $E \gappeq 20$~GeV,
 since these are affected mainly from the DM annihilations.
On the other hand,  this region is not sensitive to the solar modulation effects.
The statistical analysis,  enable us to  estimate the   $m_\mathrm{DM}$ and $\vev{\sigma v}$,
 the parameters of the DM model for the  main  annihilation channels we study.

We find that, among simple models for the annihilation final state, 
the best fit is provided by leptonic  channels, like  $\mu^+ \mu^-$ or $\tau^+ \tau^-$.
Especially the $\tau^+ \tau^-$  is  among the more
plausible final states in, e.g., supersymmetric models of DM, especially in the stau-coannihilation region.
And indeed, the positron ratio prediction for the $\tau^+ \tau^-$,  and to a less extent for the $\mu^+ \mu^-$ channel,  describes 
 the recent  AMS-02 positron ration data  strikingly well.

However, as other authors have also emphasized, the annihilation cross section
required in such a scenario is relatively large, and the stronger lower bound
on the DM particle  mass from AMS-02 compared to PAMELA, implies that the minimum
cross section must in turn be larger, many orders of magnitude larger than
what would be required for a DM candidate that was previously in
thermal equilibrium to freeze out with the cosmological DM density inferred
today. This is not in itself a problem, since one could postulate the presence of
an additional, non-thermal, origin of the rest of the requisite density.

There is, however, another important issue  with size of the required annihilation cross section.
These large cross sections may exceed the unitarity limit $(4 \pi / m_\mathrm{DM}^2) \times (2 J + 1)$,
where the initial state angular momentum $J = 0$ (1) corresponds to s(p)-wave processes. 
As we saw  this constraint is violated over a great part of the ($m_\mathrm{DM}$, $\langle \sigma v \rangle$)
 plane. In particular, only  the $\mu^+\mu^-$ and a part of $\tau^+\tau^-$ region 
 that  fit to the AMS data, survive the unitarily bound. The hadronic DM annihilation channel $b\bar{b}$ and $W^+ W^-$
 clearly vilote the unitarity limit. 
 
 Furthermore, we discussed the constraints imposed to these models from $\gamma$-ray data from Fermi-LAT and H.E.S.S
 experiments. Unfortunately, practically all the channels both the leptonic and the hadronic violate 
 the 95\% CL  bound from the Fermi-LAT related to the GC or the main dSph galaxies.
 The H.E.S.S exclusion limit is more severe and again excludes the regions of the parameter space 
 of the DM models that can explain the AMS-02 data.

Concluding we notice, that the better precision of the AMS-02 data and their
greater energy range provide stronger constraints than PAMELA on the mass and
annihilation cross section of any putative DM particle. In particular,
the AMS-02 data increase the lower limit on any such dark mass particle.
As a consequence, one can make interesting predictions for the magnitude
of the positron excess at energies $E > 350$~GeV, beyond the
energy range of the AMS-02 data released so far. Specifically, within a model
of DM annihilation to $\tau^+ \tau^-$, the AMS data restrict the rate
at which the positron fraction can diminish at energies $E > 350$~GeV.

\vspace*{6cm}
\section*{Acknowledgements}
The author acknowledges   useful comments on this work  from   John Ellis and Keith Olive.
He also  thanks CERN TH Division for its hospitality, where part of this work was done. 
This visit was supported by  London Centre for Terauniverse Studies (LCTS), 
using funding from the European Research Council via the 
Advanced Investigator Grant 267352.

\end{document}